\pdfoutput=1
\documentclass[10pt,twocolumn]{article}

\usepackage{times}
\usepackage[
    left=1.9cm,
    right=1.9cm,
    top=2.50cm,
    bottom=2.50cm
]{geometry}

\usepackage{tikz}
\usepackage{color}
\usepackage{algorithm}
\usepackage{amsmath}
\usepackage{balance}
\usepackage{subfig}
\usepackage{graphicx}
\usepackage{xspace}
\usepackage{url}
\usepackage{hyperref}
\usepackage{pifont}
\usepackage{siunitx}
\usepackage[utf8]{inputenc}
\usepackage{tabularx}
\usepackage{stfloats}
\usepackage{enumitem}
\usepackage[printonlyused,withpage]{acronym}
\usepackage[skip=6pt]{caption}
\usepackage{xspace}
\usepackage{listings}

\definecolor{CommentCol}{rgb}{0.3, 0.3, 0.3}

\lstdefinelanguage{Rust}{%
  sensitive%
, morecomment=[l]{//}%
, morecomment=[s]{/*}{*/}%
, moredelim=[s][{\itshape\color[rgb]{0.75,0,0.5}}]{\#[}{]}%
, morestring=[b]{"}%
, alsodigit={}%
, alsoother={}%
, alsoletter={!}%
, morekeywords={break, continue, else, for, if, in, loop, match, return, while}  
, morekeywords={as, const, let, move, mut, ref, static}  
, morekeywords={dyn, enum, fn, impl, Self, self, struct, trait, type, union, use, where}  
, morekeywords={crate, extern, mod, pub, super}  
, morekeywords={unsafe}  
, morekeywords={abstract, alignof, become, box, do, final, macro, offsetof, override, priv, proc, pure, sizeof, typeof, unsized, virtual, yield}  
, morekeywords=[2]{Add, AddAssign, Any, AsciiExt, AsInner, AsInnerMut, AsMut, AsRawFd, AsRawHandle, AsRawSocket, AsRef, Binary, BitAnd, BitAndAssign, Bitor, BitOr, BitOrAssign, BitXor, BitXorAssign, Borrow, BorrowMut, Boxed, BoxPlace, BufRead, BuildHasher, CastInto, CharExt, Clone, CoerceUnsized, CommandExt, Copy, Debug, DecodableFloat, Default, Deref, DerefMut, DirBuilderExt, DirEntryExt, Display, Div, DivAssign, DoubleEndedIterator, DoubleEndedSearcher, Drop, EnvKey, Eq, Error, ExactSizeIterator, ExitStatusExt, Extend, FileExt, FileTypeExt, Float, Fn, FnBox, FnMut, FnOnce, Freeze, From, FromInner, FromIterator, FromRawFd, FromRawHandle, FromRawSocket, FromStr, FullOps, FusedIterator, Generator, Hash, Hasher, Index, IndexMut, InPlace, Int, Into, IntoCow, IntoInner, IntoIterator, IntoRawFd, IntoRawHandle, IntoRawSocket, IsMinusOne, IsZero, Iterator, JoinHandleExt, LargeInt, LowerExp, LowerHex, MetadataExt, Mul, MulAssign, Neg, Not, Octal, OpenOptionsExt, Ord, OsStrExt, OsStringExt, Packet, PartialEq, PartialOrd, Pattern, PermissionsExt, Place, Placer, Pointer, Product, Put, RangeArgument, RawFloat, Read, Rem, RemAssign, Seek, Shl, ShlAssign, Shr, ShrAssign, Sized, SliceConcatExt, SliceExt, SliceIndex, Stats, Step, StrExt, Sub, SubAssign, Sum, Sync, TDynBenchFn, Terminal, Termination, ToOwned, ToSocketAddrs, ToString, Try, TryFrom, TryInto, UnicodeStr, Unsize, UpperExp, UpperHex, WideInt, Write}
, morekeywords=[2]{Send}  
, morekeywords=[3]{bool, char, f32, f64, i8, i16, i32, i64, isize, str, u8, u16, u32, u64, unit, usize, i128, u128}  
, morekeywords=[4]{Err, false, None, Ok, Some, true}  
, morekeywords=[3]{AccessError, Adddf3, AddI128, AddoI128, AddoU128, ADDRESS, ADDRESS64, addrinfo, ADDRINFOA, AddrParseError, Addsf3, AddU128, advice, aiocb, Alignment, AllocErr, AnonPipe, Answer, Arc, Args, ArgsInnerDebug, ArgsOs, Argument, Arguments, ArgumentV1, Ashldi3, Ashlti3, Ashrdi3, Ashrti3, AssertParamIsClone, AssertParamIsCopy, AssertParamIsEq, AssertUnwindSafe, AtomicBool, AtomicPtr, Attr, auxtype, auxv, BackPlace, BacktraceContext, Barrier, BarrierWaitResult, Bencher, BenchMode, BenchSamples, BinaryHeap, BinaryHeapPlace, blkcnt, blkcnt64, blksize, BOOL, boolean, BOOLEAN, BoolTrie, BorrowError, BorrowMutError, Bound, Box, bpf, BTreeMap, BTreeSet, Bucket, BucketState, Buf, BufReader, BufWriter, Builder, BuildHasherDefault, BY, BYTE, Bytes, CannotReallocInPlace, cc, Cell, Chain, CHAR, CharIndices, CharPredicateSearcher, Chars, CharSearcher, CharsError, CharSliceSearcher, CharTryFromError, Child, ChildPipes, ChildStderr, ChildStdin, ChildStdio, ChildStdout, Chunks, ChunksMut, ciovec, clock, clockid, Cloned, cmsgcred, cmsghdr, CodePoint, Color, ColorConfig, Command, CommandEnv, Component, Components, CONDITION, condvar, Condvar, CONSOLE, CONTEXT, Count, Cow, cpu, CRITICAL, CStr, CString, CStringArray, Cursor, Cycle, CycleIter, daddr, DebugList, DebugMap, DebugSet, DebugStruct, DebugTuple, Decimal, Decoded, DecodeUtf16, DecodeUtf16Error, DecodeUtf8, DefaultEnvKey, DefaultHasher, dev, device, Difference, Digit32, DIR, DirBuilder, dircookie, dirent, dirent64, DirEntry, Discriminant, DISPATCHER, Display, Divdf3, Divdi3, Divmoddi4, Divmodsi4, Divsf3, Divsi3, Divti3, dl, Dl, Dlmalloc, Dns, DnsAnswer, DnsQuery, dqblk, Drain, DrainFilter, Dtor, Duration, DwarfReader, DWORD, DWORDLONG, DynamicLibrary, Edge, EHAction, EHContext, Elf32, Elf64, Empty, EmptyBucket, EncodeUtf16, EncodeWide, Entry, EntryPlace, Enumerate, Env, epoll, errno, Error, ErrorKind, EscapeDebug, EscapeDefault, EscapeUnicode, event, Event, eventrwflags, eventtype, ExactChunks, ExactChunksMut, EXCEPTION, Excess, ExchangeHeapSingleton, exit, exitcode, ExitStatus, Failure, fd, fdflags, fdsflags, fdstat, ff, fflags, File, FILE, FileAttr, filedelta, FileDesc, FilePermissions, filesize, filestat, FILETIME, filetype, FileType, Filter, FilterMap, Fixdfdi, Fixdfsi, Fixdfti, Fixsfdi, Fixsfsi, Fixsfti, Fixunsdfdi, Fixunsdfsi, Fixunsdfti, Fixunssfdi, Fixunssfsi, Fixunssfti, Flag, FlatMap, Floatdidf, FLOATING, Floatsidf, Floatsisf, Floattidf, Floattisf, Floatundidf, Floatunsidf, Floatunsisf, Floatuntidf, Floatuntisf, flock, ForceResult, FormatSpec, Formatted, Formatter, Fp, FpCategory, fpos, fpos64, fpreg, fpregset, FPUControlWord, Frame, FromBytesWithNulError, FromUtf16Error, FromUtf8Error, FrontPlace, fsblkcnt, fsfilcnt, fsflags, fsid, fstore, fsword, FullBucket, FullBucketMut, FullDecoded, Fuse, GapThenFull, GeneratorState, gid, glob, glob64, GlobalDlmalloc, greg, group, GROUP, Guard, GUID, Handle, HANDLE, Handler, HashMap, HashSet, Heap, HINSTANCE, HMODULE, hostent, HRESULT, id, idtype, if, ifaddrs, IMAGEHLP, Immut, in, in6, Incoming, Infallible, Initializer, ino, ino64, inode, input, InsertResult, Inspect, Instant, int16, int32, int64, int8, integer, IntermediateBox, Internal, Intersection, intmax, IntoInnerError, IntoIter, IntoStringError, intptr, InvalidSequence, iovec, ip, IpAddr, ipc, Ipv4Addr, ipv6, Ipv6Addr, Ipv6MulticastScope, Iter, IterMut, itimerspec, itimerval, jail, JoinHandle, JoinPathsError, KDHELP64, kevent, kevent64, key, Key, Keys, KV, l4, LARGE, lastlog, launchpad, Layout, Lazy, lconv, Leaf, LeafOrInternal, Lines, LinesAny, LineWriter, linger, linkcount, LinkedList, load, locale, LocalKey, LocalKeyState, Location, lock, LockResult, loff, LONG, lookup, lookupflags, LookupHost, LPBOOL, LPBY, LPBYTE, LPCSTR, LPCVOID, LPCWSTR, LPDWORD, LPFILETIME, LPHANDLE, LPOVERLAPPED, LPPROCESS, LPPROGRESS, LPSECURITY, LPSTARTUPINFO, LPSTR, LPVOID, LPWCH, LPWIN32, LPWSADATA, LPWSAPROTOCOL, LPWSTR, Lshrdi3, Lshrti3, lwpid, M128A, mach, major, Map, mcontext, Metadata, Metric, MetricMap, mflags, minor, mmsghdr, Moddi3, mode, Modsi3, Modti3, MonitorMsg, MOUNT, mprot, mq, mqd, msflags, msghdr, msginfo, msglen, msgqnum, msqid, Muldf3, Mulodi4, Mulosi4, Muloti4, Mulsf3, Multi3, Mut, Mutex, MutexGuard, MyCollection, n16, NamePadding, NativeLibBoilerplate, nfds, nl, nlink, NodeRef, NoneError, NonNull, NonZero, nthreads, NulError, OccupiedEntry, off, off64, oflags, Once, OnceState, OpenOptions, Option, Options, OptRes, Ordering, OsStr, OsString, Output, OVERLAPPED, Owned, Packet, PanicInfo, Param, ParseBoolError, ParseCharError, ParseError, ParseFloatError, ParseIntError, ParseResult, Part, passwd, Path, PathBuf, PCONDITION, PCONSOLE, Peekable, PeekMut, Permissions, PhantomData, pid, Pipes, PlaceBack, PlaceFront, PLARGE, PoisonError, pollfd, PopResult, port, Position, Powidf2, Powisf2, Prefix, PrefixComponent, PrintFormat, proc, Process, PROCESS, processentry, protoent, PSRWLOCK, pthread, ptrdiff, PVECTORED, Queue, radvisory, RandomState, Range, RangeFrom, RangeFull, RangeInclusive, RangeMut, RangeTo, RangeToInclusive, RawBucket, RawFd, RawHandle, RawPthread, RawSocket, RawTable, RawVec, Rc, ReadDir, Receiver, recv, RecvError, RecvTimeoutError, ReentrantMutex, ReentrantMutexGuard, Ref, RefCell, RefMut, REPARSE, Repeat, Result, Rev, Reverse, riflags, rights, rlim, rlim64, rlimit, rlimit64, roflags, Root, RSplit, RSplitMut, RSplitN, RSplitNMut, RUNTIME, rusage, RwLock, RWLock, RwLockReadGuard, RwLockWriteGuard, sa, SafeHash, Scan, sched, scope, sdflags, SearchResult, SearchStep, SECURITY, SeekFrom, segment, Select, SelectionResult, sem, sembuf, send, Sender, SendError, servent, sf, Shared, shmatt, shmid, ShortReader, ShouldPanic, Shutdown, siflags, sigaction, SigAction, sigevent, sighandler, siginfo, Sign, signal, signalfd, SignalToken, sigset, sigval, Sink, SipHasher, SipHasher13, SipHasher24, size, SIZE, Skip, SkipWhile, Slice, SmallBoolTrie, sockaddr, SOCKADDR, sockcred, Socket, SOCKET, SocketAddr, SocketAddrV4, SocketAddrV6, socklen, speed, Splice, Split, SplitMut, SplitN, SplitNMut, SplitPaths, SplitWhitespace, spwd, SRWLOCK, ssize, stack, STACKFRAME64, StartResult, STARTUPINFO, stat, Stat, stat64, statfs, statfs64, StaticKey, statvfs, StatVfs, statvfs64, Stderr, StderrLock, StderrTerminal, Stdin, StdinLock, Stdio, StdioPipes, Stdout, StdoutLock, StdoutTerminal, StepBy, String, StripPrefixError, StrSearcher, subclockflags, Subdf3, SubI128, SuboI128, SuboU128, subrwflags, subscription, Subsf3, SubU128, Summary, suseconds, SYMBOL, SYMBOLIC, SymmetricDifference, SyncSender, sysinfo, System, SystemTime, SystemTimeError, Take, TakeWhile, tcb, tcflag, TcpListener, TcpStream, TempDir, TermInfo, TerminfoTerminal, termios, termios2, TestDesc, TestDescAndFn, TestEvent, TestFn, TestName, TestOpts, TestResult, Thread, threadattr, threadentry, ThreadId, tid, time, time64, timespec, TimeSpec, timestamp, timeval, timeval32, timezone, tm, tms, ToLowercase, ToUppercase, TraitObject, TryFromIntError, TryFromSliceError, TryIter, TryLockError, TryLockResult, TryRecvError, TrySendError, TypeId, U64x2, ucontext, ucred, Udivdi3, Udivmoddi4, Udivmodsi4, Udivmodti4, Udivsi3, Udivti3, UdpSocket, uid, UINT, uint16, uint32, uint64, uint8, uintmax, uintptr, ulflags, ULONG, ULONGLONG, Umoddi3, Umodsi3, Umodti3, UnicodeVersion, Union, Unique, UnixDatagram, UnixListener, UnixStream, Unpacked, UnsafeCell, UNWIND, UpgradeResult, useconds, user, userdata, USHORT, Utf16Encoder, Utf8Error, Utf8Lossy, Utf8LossyChunk, Utf8LossyChunksIter, utimbuf, utmp, utmpx, utsname, uuid, VacantEntry, Values, ValuesMut, VarError, Variables, Vars, VarsOs, Vec, VecDeque, vm, Void, WaitTimeoutResult, WaitToken, wchar, WCHAR, Weak, whence, WIN32, WinConsole, Windows, WindowsEnvKey, winsize, WORD, Wrapping, wrlen, WSADATA, WSAPROTOCOL, WSAPROTOCOLCHAIN, Wtf8, Wtf8Buf, Wtf8CodePoints, xsw, xucred, Zip, zx}
, morekeywords=[5]{assert!, assert_eq!, assert_ne!, cfg!, column!, compile_error!, concat!, concat_idents!, debug_assert!, debug_assert_eq!, debug_assert_ne!, env!, eprint!, eprintln!, file!, format!, format_args!, include!, include_bytes!, include_str!, line!, module_path!, option_env!, panic!, print!, println!, select!, stringify!, thread_local!, try!, unimplemented!, unreachable!, vec!, write!, writeln!}  
}%

\lstdefinestyle{colouredRust}%
{ basicstyle=\footnotesize\ttfamily%
, identifierstyle=%
, commentstyle=\color[gray]{0.4}%
, stringstyle=\color[rgb]{0, 0, 0.5}%
, keywordstyle=\bfseries
, keywordstyle=[2]\color[rgb]{0.75, 0, 0}
, keywordstyle=[3]\color[rgb]{0, 0.5, 0}
, keywordstyle=[4]\color[rgb]{0, 0.5, 0}
, keywordstyle=[5]\color[rgb]{0, 0, 0.75}
, columns=spaceflexible%
, keepspaces=true%
, showspaces=false%
, showtabs=false%
, showstringspaces=true%
}%

\lstdefinestyle{boxed}{
  style=colouredRust%
, numbers=left%
, firstnumber=auto%
, numberblanklines=true%
, frame=trbL%
, numberstyle=\tiny%
, frame=leftline%
, numbersep=7pt%
, framesep=5pt%
, framerule=10pt%
, xleftmargin=15pt%
, backgroundcolor=\color[gray]{0.97}%
, rulecolor=\color[gray]{0.90}%
}

\definecolor{StringRed}{rgb}{.637,0.082,0.082}
\definecolor{CommentGreen}{rgb}{0.0,0.55,0.3}
\definecolor{KeywordBlue}{rgb}{0.0,0.3,0.55}
\definecolor{LinkColor}{rgb}{0.55,0.0,0.3}
\definecolor{CiteColor}{rgb}{0.55,0.0,0.3}
\lstdefinestyle{small}{
  mathescape=true,
	basicstyle=\ttfamily\scriptsize,
	keywordstyle=\color{blue}\bfseries,
	identifierstyle=\color{CommentGreen},
	stringstyle=\color{StringRed},
	emphstyle=\color{LifetimeCol},
	commentstyle=\color{CommentCol}\itshape,
	tabsize=4,
        numbers=left,
        xleftmargin=14pt,
        showstringspaces=false,
        columns=flexible,
  literate=
		 {γ}{\(\gamma\)\lst@whitespacefalse}1
     {λ}{\(\lambda\)\lst@whitespacefalse}1
     {≠}{\(\neq\)\lst@whitespacefalse}1
     {∀}{\(\forall\)\lst@whitespacefalse}1
     {∃}{\(\exists\)\lst@whitespacefalse}1
     {∧}{\(\land\)\lst@whitespacefalse}1
     {∨}{\(\lor\)\lst@whitespacefalse}1
     {↔}{\(\leftrightarrow\)\lst@whitespacefalse}1
	   {⌜}{\(\ulcorner\)\lst@whitespacefalse}1
	   {⌝}{\(\urcorner\)\lst@whitespacefalse}1
     {∈}{\(\in\)\lst@whitespacefalse}1
		 {≤}{\(\le\)\lst@whitespacefalse}1
     {≥}{\(\geq\)\lst@whitespacefalse}1
		 {∀}{\(\forall\)\lst@whitespacefalse}1
		 {→}{\(\rightarrow\)\lst@whitespacefalse}1
     {π}{\(\pi\)\lst@whitespacefalse}1
     {Σ}{\(\Sigma\)\lst@whitespacefalse}1
     {□}{\(\square\)\lst@whitespacefalse}1
     {∗}{\(\ast\)\lst@whitespacefalse}1
     {ₗ}{\(_l\)\lst@whitespacefalse}1
     {◁}{\(\triangleleft\)\lst@whitespacefalse}1
}
\lstdefinestyle{footnote}{
  style=small,
	basicstyle=\ttfamily\footnotesize,
}
\newcommand{\rstinline}[1]{\mbox{\lstinline[language=rust,mathescape,style=colouredRust]{#1}}}

\newcommand{\myparagraph}[1]{\vspace{1mm} \smallskip \noindent{\bf {#1}}}
\newcommand{\nip}[1]{\vspace{1ex}\noindent\textbf{#1}}
\newcommand\ie{\textit{i.e.}}
\newcommand\eg{\textit{e.g.}}
\newcommand{\ACE} {ACE\xspace}

\newcommand{\woz}[1]{{\bf\textcolor{orange}{woz: #1 }}}
\newcommand{\mvle}[1]{{\bf\textcolor{green}{mvle: #1 }}}
\newcommand{\gdhh}[1]{{\bf\textcolor{brown}{gdhh: #1}}}
\newcommand{\erp}[1]{{\bf\textcolor{teal}{erp: #1 }}}
\newcommand{\avi}[1]{{\bf\textcolor{green}{avi: #1 }}}

\newcommand{\lennard}[1]{{\bf\textcolor{green}{lg: #1 }}}

\begin{document}
\title{ACE: Towards A High-Assurance Isolated Virtualization Environment for RISC-V}

\author{
       Wojciech Ozga\\
       \textit{IBM Research --- Zürich} \\
    \and
       Guerney D. H. Hunt\\
       \textit{IBM T.J. Watson Research Center}
    \and
       Michael V. Le\\
       \textit{IBM T.J. Watson Research Center}
    \and
       Lennard Gäher\\
       \textit{MPI-SWS, Germany}
    \and
       Avraham Shinnar\\
       \textit{IBM T.J. Watson Research Center}
    \and
       Elaine R. Palmer\\
       \textit{IBM T.J. Watson Research Center}
    \and
       Hani Jamjoom\\
       \textit{IBM T.J. Watson Research Center}
    \and
       Silvio Dragone\\
       \textit{IBM Research --- Zürich}
}



\maketitle

\begin{abstract}

Confidential computing has proven its value in cloud environments, but its potential for securing edge and high-end embedded systems (\eg{}, automotive controllers, cryptographic accelerators, telco infrastructure) remains largely unexplored. We present \ACE, an open-source, royalty-free virtualization-based confidential computing system for RISC-V targeting these environments.
\ACE isolates software into confidential virtual machines with narrow, well-defined interfaces, building exclusively on commodity RISC-V hardware without specialized hardware extensions or licensing fees.
Our prototype evaluation on the first commercially available RISC-V hardware supporting virtualization shows that \ACE is a viable candidate for our target systems.

\end{abstract}

\acresetall

\section{Introduction}
Building mission-critical systems and confirming their safety and security properties has become increasingly challenging due to growing software size, shrinking budgets, time-to-market pressures, and talent shortages. The growing frequency of cyberattacks~\cite{solarwinds, nyt_nuclear_hack, routers_critical_infrastructure_hack, dailey2021darkside, nyt_petrochemical_hack, brasil_ehealth_leak} confirms the urgency. Modern systems are composed of hundreds to thousands of software and hardware components from different vendors, making it difficult to not only protect confidential IPs but also reason about correctness of the whole system.
During verification, all interactions between components must be checked, causing verification costs to explode or even make the verification process infeasible.

Consequently, the verification phase may be shortened or eliminated to meet deadlines or cost targets, leaving undetected memory safety bugs or security backdoors in the system~\cite{nist2024xz, log4j_exploit}.
We argue that confidential computing~\cite{confidential2022technical} addresses both problems (confidentiality and verification efficacy) simultaneously. By compartmentalizing software into isolated security domains with narrow, well-defined interfaces, it confines the impact of any compromise to a single domain and reduces the scope of what must be verified to each domain individually.

Safety- and security-critical systems that must conform with regulations~\cite{en50128, nist@fips_140_3, cc@eal, rtca2011do333} can directly benefit, because these regulations mandate extensive testing and recommend or require formal methods. For example, EAL7, the most stringent Common Criteria~\cite{cc@eal} evaluation level, requires mathematical proofs of system security properties that are hard to achieve for complex systems because proof sizes grow with the square of the specification size~\cite{matichuk2015majks}, which itself grows with the codebase.
Compartmentalization  reduces this cost by limiting what must be proved per domain.

This benefit, however, only materializes if the components enforcing isolation are themselves verifiable. The challenge here is foundational: security proofs are not flat but layered, with each property depending on those below. A single instruction in a memory-unsafe language can violate memory safety and collapse every security argument built on top of it.
Achieving high assurance therefore requires a system designed from the ground up to keep each layer tractable.
We observe that a common isolation mechanism designed for verifiability, with a small, extendable core, narrow interfaces, and memory-safe implementation, can serve as a reusable foundation that different mission-critical systems build upon, inheriting its security guarantees rather than reproving them from scratch, directly reducing costs and accelerating the certification process.

Realizing this vision on RISC-V, however, requires closing a gap.
Despite rapid expansion of RISC-V into safety-critical infrastructure, \eg, automotive and  telco~\cite{omdia2024riscv,telco,vmware,sijstermans2024}, the ecosystem still lacks confidential computing technology suited for these edge and high-end embedded RISC-V systems that operate with limited power and silicon budgets.
We introduce \ACE
to close this gap.
Commercial confidential computing for x86~\cite{cheng2023intel,amd2020sev}, IBM Z~\cite{borntrager2020secure}, or IBM POWER~\cite{hunt2021pef} targets high-end server processors in data centers, while ARM's solution requires dedicated hardware and additional licensing costs~\cite{armcca}.
\ACE, by contrast, is open-source and royalty-free, and runs on commodity RISC-V hardware.
\ACE's design has contributed to the RISC-V CoVE specification~\cite{cove_spec}.
The page management subsystem (8\% of the codebase) has been formally verified for memory safety and functional correctness.

As part of our work on \ACE, we developed a set of principles and a methodology that we believe are universal and provide a basis for the design and development of firmware that requires verification.
These principles enable system designers and developers to choose a tradeoff between resources invested in verification and the obtained assurance level. For example, they might rely on strongly typed, memory-safe programming languages and their compilers for memory safety while harnessing proper encapsulation techniques with deductive verification to prove correctness of a small set of traditionally error-prone unsafe operations.
We applied these principles during the design and development of \ACE, achieving a relatively high level of trust by design, even before the full formal verification is complete.

To demonstrate maturity and functional readiness of the \ACE implementation, we ran and evaluated multi-processor Linux-based confidential virtual machines (VMs) on the first RISC-V hardware available on the market that supports virtualization. The results show that \ACE incurs low performance overhead for CPU-intensive workloads and up to 48\% overhead for multi-vCPU network-intensive workloads, results in line with commercial confidential computing technologies~\cite{misono2024confidential, yan2023performance}.
I/O performance could further be improved with dedicated hardware extensions to avoid extra copies via bounce buffers.

\nip{Our contributions:}
\begin{itemize}[topsep=0pt,leftmargin=*]
    \item Define principles and methodology to guide design of high-assurance low-level systems (\S\ref{sec:design}) and applied them to build \ACE, an open-source, royalty-free, virtualization-based confidential computing for edge and high-end embedded RISC-V systems (\S\ref{sec:design:decisions}, \S\ref{sec:impl}).
    \item Provide the first evaluation of virtualization-based confidential computing on the first commercially available RISC-V hardware supporting virtualization
    (\S\ref{sec:eval}).
    \item Extend the RISC-V CoVE specification with the \ACE's unique design points, such as single-step TVM promotion, static memory partitioning, and local attestation. We extend the Linux CoVE patches to support these features.
\end{itemize}

\section{Overview of \ACE}
\label{sec:design_meth}

\ACE brings virtual machine (VM)-based confidential computing capabilities to edge and high-end embedded RISC-V systems. Like in other such architectures, a hypervisor manages the VM's lifecycle but is not part of the trusted computing base (TCB). \ACE utilizes firmware and hardware to build a trusted execution environment (TEE)~\cite{GlobalPlatlformPPv1.3}.

\begin{figure}[tbp!]
    \centering
    \includegraphics[width=0.46\textwidth]{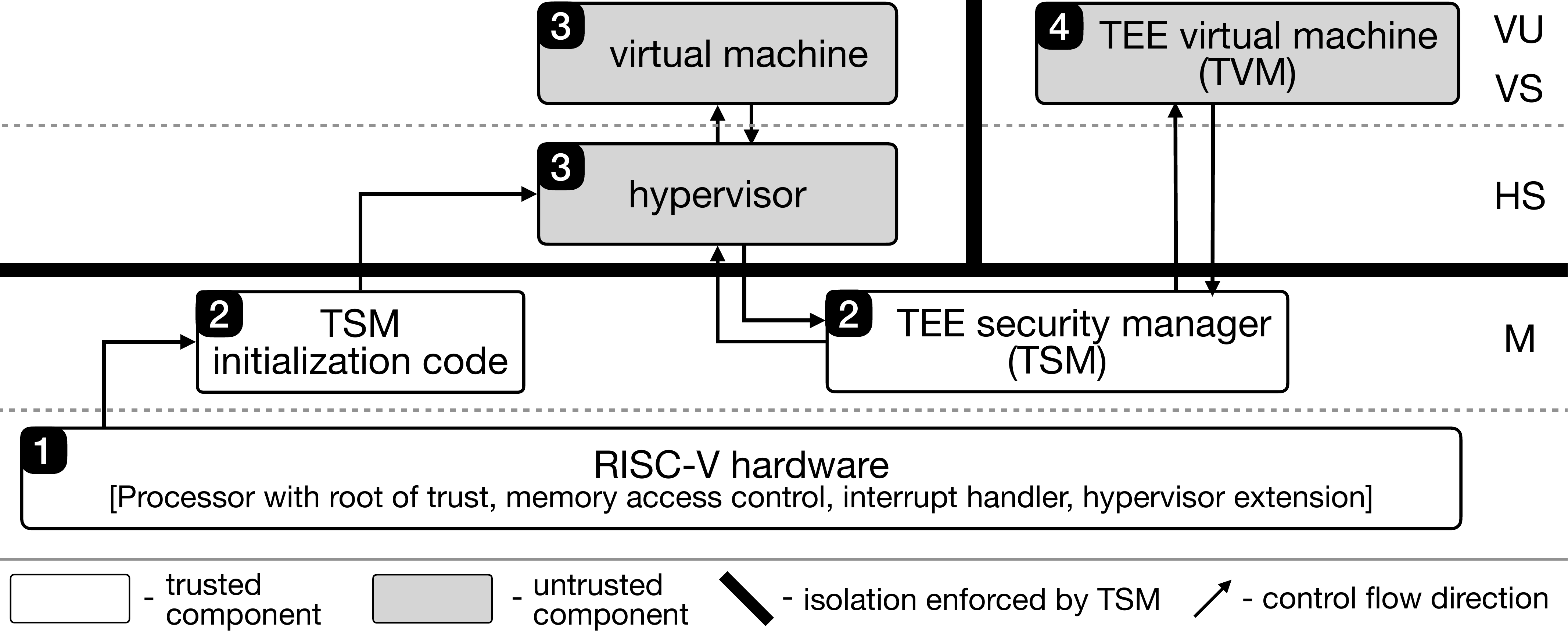}
    \caption{High-level overview of \ACE, a VM-based confidential computing targeting RISC-V.
    The TSM (\raisebox{-1pt}{\ding{203}}) leverages hardware features (\raisebox{-1pt}{\ding{202}}) to multiplex execution of different security domains (\raisebox{-1pt}{\ding{204}}) and (\raisebox{-1pt}{\ding{205}}) on top of the same hardware while preserving security guarantees.
    }
    \label{fig:overview}
\end{figure}

\label{sec:design:overview}
\autoref{fig:overview} shows a high-level overview of the
architecture, which consists of four components:
(\raisebox{-1pt}{\ding{202}}) hardware,
(\raisebox{-1pt}{\ding{203}}) TEE security manager (TSM),
(\raisebox{-1pt}{\ding{204}}) hypervisor with virtual machines (VMs),
and (\raisebox{-1pt}{\ding{205}}) TEE virtual machines (TVMs).\footnote{We follow the nomenclature used in the RISC-V CoVE~\cite{cove_spec} and TDISP~\cite{tdisp} specifications. In the literature, one can also find the terms CVMs, SVMs, or enclaves, as well as security monitor, security manager, or isolation monitor.}
The TSM's goal is to isolate security domains from each other.
The hypervisor and normal VMs constitute one security domain and each TVM constitutes a distinct security domain.
The TSM is a piece of code that switches execution of a security domain's (a TVM or the hypervisor) context and ensures that only allowed information is exchanged.
It achieves this by reconfiguring the hardware to enable proper access controls and clearing the execution state so that there are no execution traces left when another security domain resumes.

To maintain full control over the hardware's state and configuration, the TSM asserts control of the system during the early boot process and retains its privileged role during the system's lifetime. This role allows it to enforce access controls  by switching processor privilege modes, applying memory access controls, and taking control over interrupts.

\subsection{Open Source and Compliance}
\ACE complies with the RISC-V CoVE specification~\cite{cove_spec} and builds on top of the canonical architecture~\cite{ozga2023canonical}, which specifies the minimal set of features needed for a confidential computing system. The simplicity of this design aims to streamline the formal verification process, while compliance with standards should facilitate adoption.
The project's open-source nature encourages community participation, leading to extensions that improve use-case-specific performance.

\subsection{Threat Model}\label{sec:design:threat_model}
We assume a software-level adversary whose goal is to compromise a victim's TVM, specifically (1) by tampering with the TVM's execution integrity by modifying its control or data flow execution, (2) by reading the confidential TVM's data, or (3) by impersonating the victim's TVM to its owner.
\ACE's security goal is that the adversary learns nothing about the TVM's state beyond what passes through an explicitly reclassified channel.

We assume an adversary who has root access to the hypervisor and might run her own TVMs concurrently and in parallel with victim's TVMs. She can also control the lifecycle of victim's TVMs and input data via hypervisor-controlled registers, shared memory buffers, virtual I/O devices, or the initial TVM's image state and the flattened device tree (FDT).

We exclude micro-architectural side-channel attacks~\cite{Kocher2018spectre, foreshadow} because these attacks require separate hardware mitigations such as speculation barriers and cache partitioning, which are orthogonal to \ACE's isolation mechanisms and can be applied independently.
We also exclude from the threat model physical attacks on memory, buses, and processors, because of the existence of well-known counter-measures, \eg{}, fault-detection mechanisms like lockstep, error correction code-protected memory, or encrypted buses.
These technologies should be implemented separately for target systems and do not require changes to the TSM.

\section{Design} \label{sec:design}

The design principles outlined below arise from the need to satisfy three competing constraints: resource limitations of edge and high-end embedded systems, complex security needs of confidential computing, and tractability of formal verification.

\nip{Edge and high-end embedded systems} have limited power and silicon budget and are designed to execute specific workloads with predictable runtime requirements. They may or may not be network-connected. In contrast to low-end embedded systems, \eg, sensors, our target systems contain multiple security domains and use hardware-based isolation mechanisms.
Examples are: (1) Automotive systems~\cite{redhat}, software-defined vehicles~\cite{bosch}, (2) PCIe crypto and AI accelerators~\cite{nvidia_gpu}, and (3) Telco infrastructure~\cite{telco,vmware}.

\nip{Confidential computing}, which enables isolation of security domains on the same hardware, builds on hardware and firmware components whose implementation and composition require rigorous security analysis.
While increased complexity of these components might enhance the performance and capabilities of target systems, it also poses greater challenges for ensuring the desired security.

\nip{Formal verification} scales best when the verification can be modularized with clear abstraction boundaries, making reasoning about system invariants local.
More complex representation invariants (\eg{}, due to low-level memory hacks) complicate reasoning.

Achieving \ACE's goal while satisfying the above constraints requires tradeoffs.
We define the following guiding principles to
navigate these tradeoffs effectively and justify the design decisions described in \S\ref{sec:design:decisions}.

\begin{itemize}
    \item[P1:] Design a small modular core that targets a simple, well-defined hardware baseline, while remaining extensible to more complex use cases and hardware.
    \item[P2:] Minimize cross-domain interactions to make formal verification achievable.
    \item[P3:] Encapsulate complex operations with clearly defined boundaries (modularity, extensibility to satisfy formal verification requirements).
    \item[P4:] Treat verification as a first-class citizen, designing and implementing with verification in the loop, using verification considerations to guide the system design to reduce verification time and improve coverage.
\end{itemize}

\begin{figure}[tbp!]
    \centering
    \includegraphics[width=0.46\textwidth]{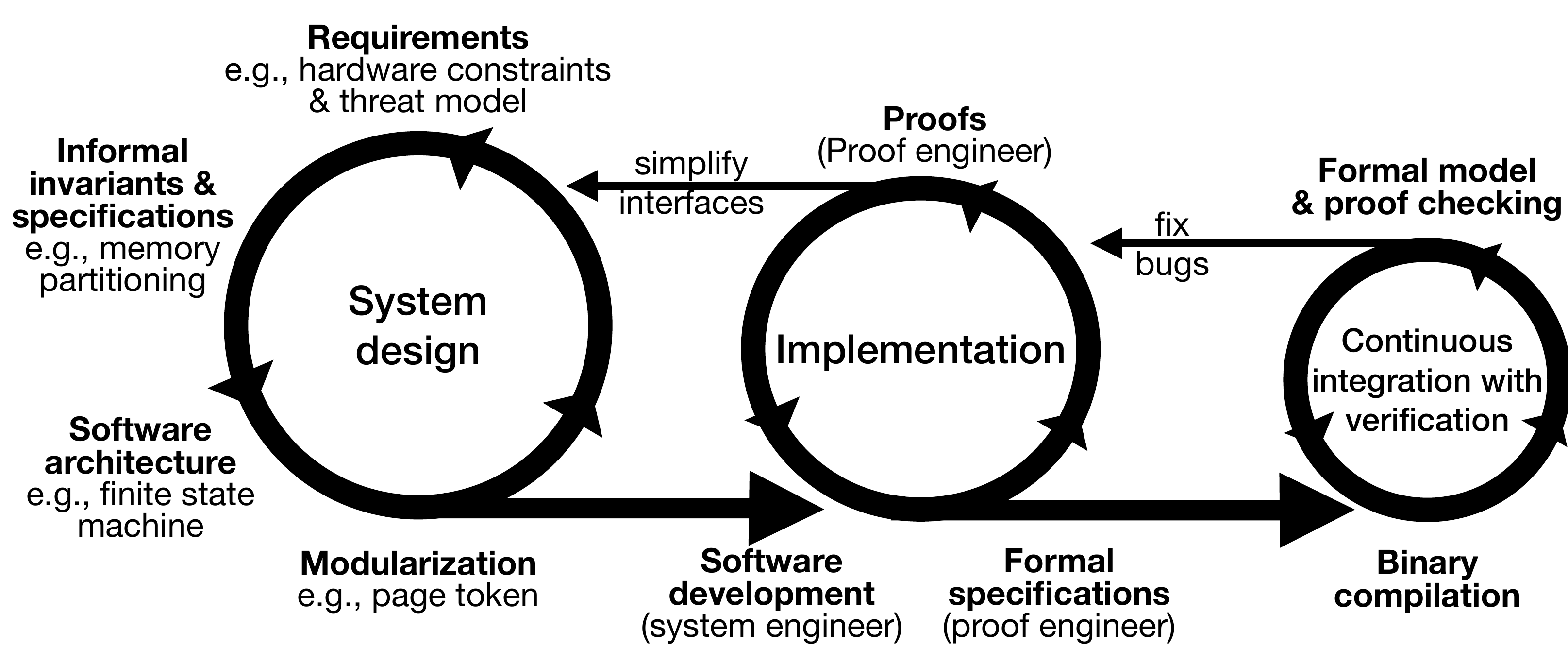}
    \caption{Our methodology to develop formally verified low-level computing systems.}
    \label{fig:workflow}
\end{figure}

\subsection{Design Methodology}
\label{sec:design_methodology}
\autoref{fig:workflow} shows our methodology for designing and developing \ACE, consisting of three phases interacting with each other: system design, implementation, and continuous integration with proofs.
For the remainder of this section we will focus on the system design decisions, the other two phases in our process will be discussed in \autoref{sec:impl}.

We started our design phase by evaluating the requirements such as hardware constraints and the threat model (\autoref{sec:design:threat_model}).
Those requirements, together with our design principles, then motivated our key design decisions and high-level invariants (\autoref{sec:design:decisions}).

\subsection{Key Design Choices \& Invariants}
\label{sec:design:decisions}
By adhering to our design principles and applying our methodology, we made design decisions that distinguish \ACE: (1) prefer commonly used, simple hardware primitives with easily abstracted functionality, (2) favor static versus dynamic configurations for simplified reasoning about system state, (3) avoid unnecessarily complex interfaces to reduce intermediate states, (4) prioritize simplicity over performance and memory usage to facilitate formal verification.

\myparagraph{Simple hardware and abstractions.}
Since \ACE primarily targets edge and high-end embedded systems, its design emphasizes compatibility with widely available, standardized hardware components for broad adoption.
Following principle \textbf{P1}, we do not require the presence of specialized TEE-optimized hardware elements, like sophisticated memory isolation, interrupt delivery, or IO mapping mechanisms, typical for high-end application servers, like in Intel TDX, AMD SEV, or ARM CCA.
Instead, our goal is to build on simple, general-purpose hardware mechanisms \cite{lee2020keystone, ferraiuolo2017komodo, ozga2023canonical}, while allowing additional hardware and software extensions to enhance performance of use-case-specific functionality when scaling from high-end embedded to edge or beyond.

We lay \ACE's design on a virtualization layer because it provides a good abstraction over the execution environment~(\textbf{P3}) and is a natural boundary for isolation with a smaller attack surface~(\textbf{P2}) compared with process-based TEEs~\cite{costan2016sanctum, lee2020keystone, sgx}. We expect our targeted systems to be powerful enough to support a RISC-V processor with virtualization capabilities, i.e., the hypervisor extension and a two-level address translation.\footnote{The RISC-V hypervisor extension can be emulated in firmware~\cite{waterman2021riscvpriv}, allowing \ACE to run on simpler hardware at the cost of reduced performance.}
These assumptions are met by today's operational technology (OT) systems~\cite{abb_ot, ot_virt}, automotive-grade SoCs~\cite{redhat}, AI accelerators~\cite{nvidia_gpu}, and cryptographic coprocessors~\cite{marvell}.
Systems with requirements for better I/O performance, memory utilization, or lower power consumption can extend hardware with dynamical memory partitioning~\cite{sahita2023smmtt}, directly injected interrupts~\cite{riscv_aia}, and directly assigned IO devices~\cite{coveio, tdisp}.

While the concept of \ACE can also be applied to high-end application servers (not restricted to only edge/embedded systems), we think it is less practical and effective given the larger TCBs and hardware complexity needed to support the wide class of workloads demanded by such systems.

\myparagraph{Statically partitioned confidential memory.}
The TSM uses memory access control hardware to prevent unauthorized accesses to protected memory regions.
Embedded systems usually run a fixed well-understood set of applications. This makes them uniquely suited to static memory partitioning because the memory requirements are predictable.

There are different hardware-based mechanisms used for memory access control. More complex mechanisms enable fine (page-level) access control at the cost of more complex configuration setup and runtime walks. Alternatively, simpler mechanisms generally have a limited number of coarse-grained access control rules for physical memory but are simple to set up, implement, and consequently verify.

Following \textbf{P1} and \textbf{P2}, \ACE statically partitions memory into non-confidential and confidential memory. This one-time operation utilizes simple hardware that performs address range checking, simplifying the mathematical modeling and verification of the system due to easier ownership reasoning. This advantage comes at the cost of potential memory over- or under-allocation, which is acceptable for our target systems. Their resource requirements and applications are known upfront, making static allocation practical.

\myparagraph{Single-step TVM creation.}
\ACE requires only a single hypervisor call to instantiate a TVM. The hypervisor first creates a standard VM in memory from a TVM image and then requests the TSM to \emph{promote} it to a TVM. This design adheres to principles \textbf{P2} and \textbf{P3}, offering advantages over the multi-step procedures employed by existing confidential computing frameworks. In particular, the attack surface is reduced because:
(1) The adversary cannot influence the physical allocation of TVM pages, a vulnerability previously exploited to compromise confidentiality~\cite{batteringramsp26,tee-fail}.
(2) There is no intermediate or partially initialized TVM state that could be targeted for rollback, an attack analogous to a TPM reset attack.
(3) The absence of multiple calls eliminates the risk of reaching an unsafe state with specific combinations of request order and parameter values.

Since the TSM execution is non-interruptible, the hardware thread (a hart in RISC-V terminology) remains blocked while the TSM copies the VM into confidential memory. This overhead is generally acceptable because most TVM memory consists of zeroed pages, which are skipped during promotion and lazily mapped at runtime.

\begin{figure}[tbp!]
    \centering
    \includegraphics[width=0.46\textwidth]{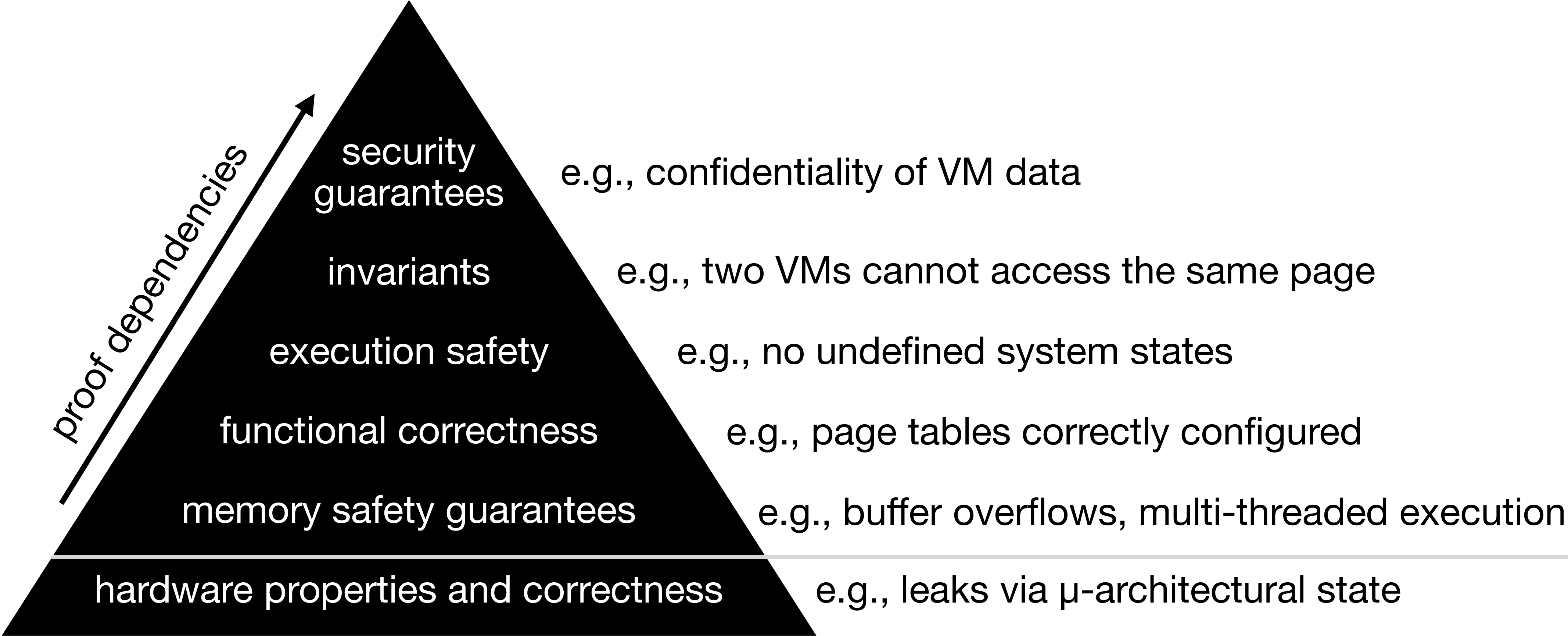}
    \caption{Pyramid of proof dependencies.}
    \label{fig:pyramide}
\end{figure}

\myparagraph{Formal Verification and Rust.}
\label{sec:formal_verification}
Our target systems might operate in regulated industries that require formal proofs.
\autoref{fig:pyramide} shows the hierarchy of properties the TSM must establish in order to achieve the security goal stated in \S\ref{sec:design:threat_model}. These are presented in the form of a pyramid: to reason about properties in one layer, one must first prove the properties in layers below.
At the base lies hardware correctness and memory safety, necessary preconditions for any higher-level argument.
Above those sit functional correctness and execution safety, and at the top sits the security guarantee: a TVM's execution state and memory contents remain confidential and unmodified except through explicitly declassified channels.
For the top-level security property,
we target a relaxed version of non-interference which accounts for the intentional information flows that a strict non-interference property would not permit.

We assume hardware correctness~\cite{stella2024enclaves, ceesay@ucfi} and focus on verification of higher layers.
Memory safety errors continue to make up a large part of security-related bugs~\cite{chromium_safety, android_safety}.
Thus, following \textbf{P4}, we decided to drastically reduce the potential surface for memory safety errors by using Rust~\cite{matsakis2014rust} as the TSM implementation language.
Rust has a rich type system that statically eliminates memory-safety issues for the vast majority of the code (written in safe Rust) without getting formal verification engineers involved, thus shifting a significant part of the verification burden to the compiler.
By applying software engineering practices such as encapsulating unsafe Rust and creating well-defined abstractions with the help of modularization (\textbf{P3}), developers gain greater confidence in the program's correctness, even in multi-threaded environments.

While ensuring memory safety eliminates many common bugs, it is just a step towards the top-level security property in \S\ref{sec:design:threat_model}.
To formally state such a hyperproperty and mechanically prove it we need the high expressiveness offered by deductive verification frameworks like Rocq~\cite{coq}.
We use the RefinedRust~\cite{gaeher2023refinedrust} framework to prove properties about Rust code in the Rocq prover and continuously check proofs for every code change in a continuous integration (CI) system.

\myparagraph{Local attestation.}
It is essential for \ACE to detect unauthorized modifications to the firmware, the TSM, or a TVM.
Because embedded systems might operate in environments with limited or no network access (\eg{}, controllers used in operational technology or automobile industry),
\ACE employs local attestation~\cite{hunt2021pef}.
This mechanism relies on information bundled with a TVM's image to verify the TVM's integrity and authenticity at creation time.
A TVM owner uses an encapsulation key of a target device to cryptographically protect a TVM attestation payload (TAP), which is a file that stores TVM integrity measurements and TVM-specific secrets. Only the expected TSM running on the correct hardware can read the contents of a TAP by accessing the decapsulation key.
Following \textbf{P1}, the presence of local attestation does not exclude remote attestation for systems with network connectivity.

\section{Implementation}
\label{sec:impl}
We implemented the TSM in Rust~\cite{matsakis2014rust} and open sourced it. 
The TSM is statically linked against the OpenSBI~\cite{opensbi} firmware to which it delegates some requests, like handling unaligned memory accesses or inter-processor interrupts (IPIs). Both OpenSBI and the TSM run in the most-privileged RISC-V M mode. OpenSBI is used as a library to simplify its future replacement or de-privileging as reported by Castes et al.~\cite{castes2024miralis}.
We use Linux KVM and QEMU as the hypervisor.
We patched the Linux kernel with the CoVE patches, which we co-developed with the RISC-V community.

The TSM's core implementation consists of: (1) initialization code executing during secure boot (\S\ref{sec:impl:init}) and (2) runtime code that drives a finite state machine (\S\ref{sec:impl:fsm}), which reacts to interrupts.
In (\S\ref{sec:formal}), we discuss how formal verification is imbued in the TSM's design and implementation.

\subsection{System Initialization}
\label{sec:impl:init}
At power-on, the hardware root of trust initializes the secure boot process~\cite{arbaugh1997secureboot}, and transfers control to the TSM initialization code. This code initiates the device identifier composition engine (DICE) chain and locks access to the unique device secret. It also configures the system such that the TSM maintains full control until the next power-cycle by reconfiguring hardware to protect the TSM's code, data, and hardware configuration from tampering by less-privileged software. Finally, it also ensures that execution of less-privileged software traps to the TSM at well-defined entry points.

\myparagraph{Memory partitioning.}
During the system initialization, the TSM divides main memory into two contiguous regions:  non-confidential memory and confidential memory.\footnote{A TSM supporting RISC-V Smmtt might enable a hypervisor to dynamically convert non-confidential memory to confidential memory at runtime.}
Later, at runtime, the hypervisor will have complete ownership of the non-confidential memory but will not be able to access the confidential memory.
That allows the TSM to store security-sensitive data required to maintain confidential computing functionality in confidential memory, for instance the TSM's stack areas (for every physical hart), the TSM's heap area, save-state areas (for every physical and virtual hart), and code and data of TVMs.

To enforce memory access protection to confidential memory, the TSM configures hardware-based memory access control mechanisms. Specifically, it uses RISC-V physical memory protection (PMP)~\cite{huyghebaert2023formalizing} and IOPMP~\cite{ku2023iopmp}.\footnote{The current implementation of \ACE runs on a RISC-V evaluation board that does not support IOPMP because IOPMP had not been ratified yet at the time when the board was designed.}
The RISC-V architecture guarantees that only software running in M-mode is authorized to reconfigure it.

\myparagraph{Page allocator and page tokens.}
Confidential memory is further partitioned into pages that the TSM assigns to TVMs to store their code and data, as well as pages to manage the TSM's internal data structures. To ensure correct allocation of pages, \ie, that pages belong to confidential memory and have a single owner, the TSM instantiates a component called the \emph{page allocator}. The page allocator is a software submodule of the TSM. During system initialization, for every confidential memory region corresponding to a physical page, it creates a logical \emph{page token}, a Rust object that lives on the TSM's heap. Initially, the page allocator owns all page tokens. Whenever the TSM allocates a new page of memory, for example to create a TVM's data page, the TSM retrieves a free page token from the page allocator and stores it within the Rust data type corresponding to the final entity, \eg, a page table associated with the TVM. In the above example, the ownership transfer would correspond to the move of the page token's \emph{self} reference from the page allocator to the object of the TVM's page table type. Rust's type system ensures that a page token is not spent twice. In the above example, it would statically ensure that every physical memory region is only assigned to one TVM.

\subsection{Runtime}
\label{sec:impl:fsm}
The firmware executing the secure boot eventually transfers control to the hypervisor, marking the transition of the device into its operational state---\emph{runtime}. During runtime, the hypervisor, running in the HS-mode, manages which software runs on the processor and for how long.
The hypervisor has limited privileges and requests the TSM to perform security-sensitive operations, for example, accessing certain memory-mapped input/output (MMIO) or running TVMs.

To limit the attack surface, the TSM exposes only a narrow ABI to the hypervisor. In addition to the standardized RISC-V ABI calls, such as read-only calls to discover the system configuration and a call to shutdown/reboot the system, the TSM adds three calls to manage a TVM: promote, run, and destroy. By contrast, a full RISC-V CoVE defines sixteen additional complex calls for page table management and two extra calls for constructing a TVM in multiple steps.
Following \textbf{P1}, the TSM supports optional symmetrical multiprocessing calls (like start/stop/suspend hart, execute remote fence), access to hardware devices via MMIO, and emulated access to misaligned load/stores.

\begin{figure}[tbp!]
    \centering
    \includegraphics[width=0.48\textwidth]{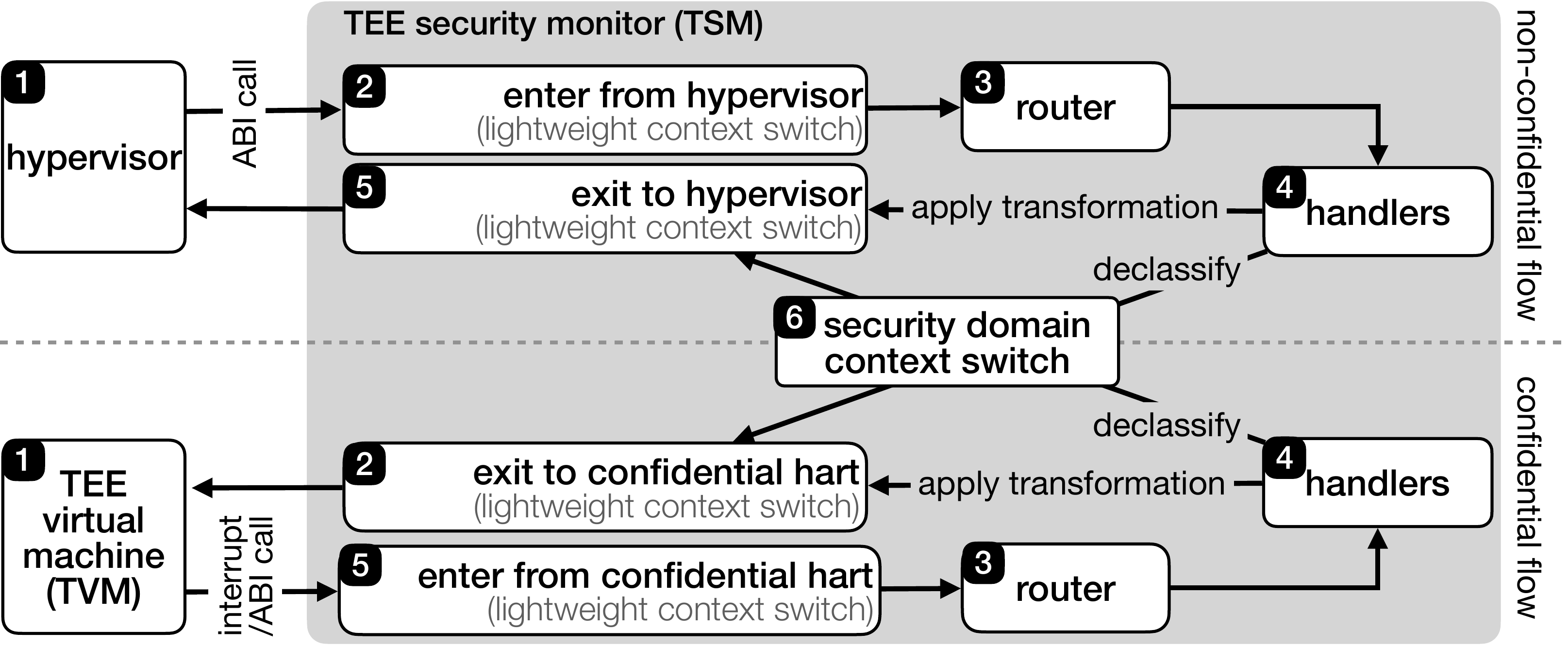}
    \caption{Finite state machine (FSM) showing the execution flow of the TSM on a single physical core on which the hypervisor and a TVM execute concurrently. }
    \label{fig:fsm}
\end{figure}

\myparagraph{Finite state machine.}
To enhance the correctness of the TSM’s implementation and simplify the verification process, the TSM operates as a finite state machine (FSM), as shown in \autoref{fig:fsm}. The hart that the TSM executes upon taking control is in one of two states: executing a non-confidential flow (upper part) or executing a confidential flow (bottom part). This split simplifies the analysis of information flow between the hypervisor and the TVM, while minimizing the potential for programming errors. We leverage Rust's modularization capabilities to statically ensure that the TSM's functionality is accessible exclusively to the hypervisor or a TVM.

In the non-confidential flow (upper part of the FSM), the TSM handles requests on behalf of the hypervisor and the hardware remains configured to enforce hypervisor-specific access control permissions. For example, the memory access control mechanism enforces that the hypervisor can access only non-confidential memory. Therefore, no additional security measures must be taken to return execution to the hypervisor.
Analogously, in the confidential flow (bottom part), the TSM processes requests of a TVM and the hardware is configured to enforce access permissions of a TVM. When the TVM resumes execution, the memory management unit (MMU) is using the page tables associated with the TVM which allows the TVM to access only memory pages it owns. If there is a need to switch from the current TVM, the TSM switches execution to the hypervisor as the hypervisor retains scheduling responsibility.

\myparagraph{Lightweight context switch.}
The TSM executes only in response to interrupts and ABI calls from the hypervisor and TVMs (\raisebox{-1pt}{\ding{202}}). The lightweight context switch (\raisebox{-1pt}{\ding{203}}) starts by storing the minimal set of the architectural state that the TSM could overwrite during its execution (\ie, general-purpose registers (GPRs) and control status registers (CSRs)) in the save-state area in main memory. Then, the router component (\raisebox{-1pt}{\ding{204}}) parses the request and invokes the appropriate handler. The handler (\raisebox{-1pt}{\ding{205}}) decides if the request will return to the same security domain or a different security domain. Handlers implement the logic that decides how to transform the architectural state to fulfill the request.
This transformation is then applied to the hardware architectural state and/or state in the save-state area. If the security domain will not be changed, the lightweight context switch (\raisebox{-1pt}{\ding{206}}) is called to return to the security domain.

\myparagraph{Security domain context switch.}
One security domain can request assistance from another security domain to perform an operation. For example, a TVM that cannot access a physical network card directly uses the VirtIO protocol, so that the hypervisor emulates such access.
The corresponding handler (\raisebox{-1pt}{\ding{205}}) then routes the call to the \emph{security domain context switch} (\raisebox{-1pt}{\ding{207}}) and the TSM crosses the boundary between the non-confidential and confidential security domains.

The security domain context switch is a piece of code that reconfigures the hardware to prepare for execution of a different security domain. It saves all architectural state of one security domain in the save-state area in the main memory and loads the architectural state of the other security domain from the main memory into hardware. All state for this context switch is stored to or read from confidential memory. The security domain context switch also reconfigures the memory access control mechanism to protect the current security domain from the one that will execute next, and clears micro-architectural caches, \eg, translation lookaside buffers (TLBs).
The Rust implementation ensures that every execution path between non-confidential and confidential flows goes through this context switch.

\myparagraph{Handlers and Reclassification.}
Handlers implement the TSM's functionality exposed to the hypervisor and TVMs via the ABI.
To reduce the risk of implementation errors creating security bugs, each handler operates only on a subset of the architectural state of a security domain. For example, a TVM's attestation handler will never have write access to the hypervisor's state, so that it can never leak attestation-related data to the hypervisor.
To do so, we leverage Rust's encapsulation to ensure a limited visibility scope and privacy of data structures in the implementations of handlers. Specifically, each handler consists of three phases: a \emph{constructor}, a \emph{transformation method}, and a \emph{destructor}.
The \emph{constructor} has read-only access to the hart state, which permits it to read information required to process the request.
The \emph{transformation method} has access only to the TSM's core functionality, such as the page allocator, the interrupt controller or the TVM's metadata storage. Finally, the \emph{destructor} has write access to the target security domain's architectural state to which it can apply or reclassify processed information.
The Rust-enforced constraints placed on handlers and accessibility of each domain's architectural state simplify how a TVM's owners and verification tools can certify that the TSM correctly implements the specification.

Importantly, it is the responsibility of a security domain to use handlers securely.
For example, for VirtIO, the TSM enables a TVM to have a shared page with the hypervisor, but does not enforce any form of data secrecy. Thus, to maintain security, a TVM itself must encrypt data before sending them to disk over VirtIO and decrypt when reading back.

\subsection{Selected Non-confidential Flow Handlers}
\myparagraph{Promote to TVM.}
As explained in \autoref{sec:design:decisions}, the TSM implements a novel single-step TVM creation procedure in which it atomically converts an instance of a VM into a TVM as part of the process called \emph{promotion}.
In contrast to other architectures~\cite{cheng2023intel, hunt2021pef, borntrager2020secure} which convert the VM memory page-by-page,
there are no intermediate states corresponding to a TVM that could be exploited by an adversary~\cite{batteringramsp26,tee-fail}.

As part of the promote request, the TSM receives as input the initial state of the VM: the registers and a program counter for the boot hart, a pointer to the root page table, a pointer to the flattened device tree, and a pointer to the attestation payload. It then creates save-state areas for the new confidential harts and traverses the VM's page table hierarchy copying non-zeroed pages from non-confidential memory into confidential memory. After measuring the initial TVM state, the TSM performs local attestation to authorize the TVM's execution.
A running TVM retrieves secrets from the TSM, such as a Linux unified key setup (LUKS) key to decrypt the root filesystem, using a dedicated ABI call.

\label{sec:imp:attest}
\myparagraph{Local attestation.}
Local attestation provides attestation for TVMs that execute with limited network access.
When the hypervisor requests the TSM to create a TVM, it passes to the TSM both the TVM's image and a file called TVM attestation payload (TAP). The TSM verifies the TVM's integrity by comparing the TVM's measurements with reference measurements from the TAP and stores the TVM's secrets retrieved from the TAP in confidential memory. A verified TVM uses a dedicated ABI call to retrieve secrets from the TSM.

The TVM's integrity measurement includes the cryptographic hashes of the TVM's code and data, flattened device tree, and the initial boot hart's state. This is stored in dedicated registers analogous to platform configuration registers (PCRs)~\cite{will2015tpm}. The TVM's data integrity is a single hash, the value of which uniquely represents the initial content of the TVM's code and data pages. 

The TVM's owner creates a TAP by encrypting a payload, which contains integrity measurements and secrets, and concatenates it with a \emph{lockbox}. A lockbox stores the symmetric key encrypted with the public portion of the TSM's attestation key, which is an asymmetric cryptographic key associated with the hardware and derived with DICE~\cite{tcg2020dice}, as defined in CoVE~\cite{cove_spec}. Only the TSM running in the target hardware can decrypt the symmetric key and payload.

The TAP supports post-quantum cryptography (PQC), so that an adversary possessing a powerful enough quantum computer is not able to decrypt the TAP.
Specifically, the TSM uses quantum-safe algorithms such as ML-KEM~\cite{fips203} to encapsulate the TAP symmetric key in the lockbox, SHA-384 to calculate the TVM's integrity measurements, and AES-GCM-256 to encrypt the payload.
The TAP also supports cryptographic agility, because each lockbox defines the type of algorithm that encrypts the TAP symmetric key to allow a TVM owner to choose a PQC algorithm of their choice.

\subsection{Selected Confidential Flow Handlers}
\myparagraph{Symmetrical multiprocessing.}
\ACE supports symmetrical multiprocessing which enables a TVM to have multiple confidential harts.
When creating a TVM, the TSM enables only the TVM's boot hart and sets other confidential harts to the powered off state. The boot hart can then request the TSM to start another confidential hart at the indicated guest physical address and with specific initial arguments.
The TSM follows the finite state machine defined in the hart state management (HSM) extension to the RISC-V supervisor binary interface (SBI)~\cite{riscv_sbi_v2} to track and ensure proper lifecycle state changes.
The TSM also permits confidential harts to send inter processor interrupts (IPIs), as well as dedicated commands to clear remote caches, to other confidential harts as defined in the respective SBI extensions~\cite{riscv_sbi_v2}.

\myparagraph{Timer programming and interrupt handling.}
The \ACE implementation relies on the hypervisor to manage a TVM's external interrupts and on hardware to provide the per-hart timers.
\ACE also supports systems with a basic hardware interrupt controller~\cite{riscv_plic} without requiring support for the more complex RISC-V AIA specification~\cite{riscv_aia}.

During the TVM's execution, all of the guest's interrupts (RISC-V VS-mode) trap directly into the TVM, owing to the TSM's interrupt delegation setup. Other interrupts (\eg{}, external or software interrupts) trap into the TSM. For external interrupts, the TSM returns execution to the hypervisor so that it can delegate to the recipient of the interrupt. If the interrupt targets a TVM, the hypervisor requests the TSM to resume the TVM's execution and inject the interrupt whose identifier is in a dedicated CSR. After the security domain context switch, the TSM decides whether to inject the interrupt to a TVM based on the interrupts the TVM agreed to receive.
If the TSM is not executing, all interrupts go either to the hypervisor or the VM as indicated by the configuration.

A TVM manages its own timer by programming a dedicated CSR specified by the RISC-V Sstc extension~\cite{waterman2021riscvpriv}. The TSM ensures proper resumption of the timer after the security domain context switch. Following the CoVE spec, the TSM discloses the timer value to the hypervisor for scheduling purposes.

\subsection{Formal Verification}
\label{sec:formal}
We have designed \ACE with the goal of providing a high-trust implementation in mind. This influenced the design of the ABI, as well as its implementation, where unsafe code is minimized and safely encapsulated.

\nip{Rust memory safety.}
Rust aims to provide memory safety guarantees, which hold as long as only the safe fragment of the language is used.
However, for low-level software like the TSM, \emph{unsafe} code is inherently necessary to interface with the hardware and do low-level memory manipulation.
Unfortunately, a memory safety error in a piece of unsafe code can nullify the memory safety guarantees of all other code, including safe Rust code.
Thus, it is important to both \emph{minimize} the amount of unsafe code and to scrutinize it.

The TSM implementation minimizes unsafe code by developing \emph{safe abstractions} over core sources of unsafety that are re-used across the code. One such abstraction are the page tokens mentioned earlier.
In total, the TSM codebase has 55 unsafe blocks, each of which is usually very short, with an average size of just over 4 lines.
Of these unsafe blocks, 12 are part of the TSM's core memory management, 11 are for interfacing with the hardware (\eg{} reading CSRs), and 20 for flow handling and managing the TVM's state.
In total, about 240 of the 8000 lines of Rust code are unsafe; about 90 of the unsafe lines are for inline assembly backing up or modifying registers.
Thus, we argue that effective low-level systems programming is possible with relatively little unsafe code.

\nip{ABI design.}
The primary attack surface of the TSM is the ABI. For that reason, in our design we minimized the ABI between the TSM and hypervisor, and between the TSM and a TVM. The \emph{promote} call is the most complex ABI with the largest input. To handle this call, the TSM must traverse the multi-level page table hierarchy and parse individual intermediate page tables. Because the hypervisor creates these page tables, it can maliciously craft the content of these page tables to exploit a buggy TSM implementation.

For example, one of the attacks consists of defining a pointer to a data page that resides in the confidential memory. During a TVM promotion, an incorrectly implemented TSM would copy the contents of an arbitrary confidential memory region to a new confidential page mapped to the attacker's TVM. To mitigate this class of attacks, the TSM allows a hypervisor and a TVM to pass as input arguments only memory addresses they own, \ie, non-confidential memory and guest physical addresses, respectively.
To enforce this requirement statically during compile time, we define dedicated Rust types for each type of memory address.
At ABI entry points, all memory addresses are validated in order to obtain valid (non-)confidential memory addresses that can be trusted in the rest of the security monitor.
Such protections are possible thanks to Rust's rich type system which guarantees non-trivial properties even before applying complex deductive verification tools.

\nip{Methodology.}
We integrated formal verification methodology into our design and implementation using {RefinedRust}~\cite{gaeher2023refinedrust}.
RefinedRust is a modular verification tool in which individual functions are given a specification and are verified independently of each other to then compose to a verification result for the whole program.
RefinedRust verifies functional correctness of the code against the specification, and memory safety of safe and unsafe code as well as absence of panics.

During the design phase, we stated individual module invariants that the implementation has to uphold.
After the module implementation, we annotated the Rust code with specifications for individual functions and data structures.
If it turned out that writing the specifications for interfaces was overly complex, the proof engineer and system engineer iterated to simplify the code, making it more amenable to formal specification.

RefinedRust generates a formal model of the Rust code and specifications in the Rocq prover as well as a proof template using RefinedRust's automation tactics.
If the automatic proof does not succeed, the proof engineer interactively adds hints to the proof template.
The specification and proof are then checked into continuous verification, where RefinedRust continuously updates the model of the changing code, and checks the specifications and proofs against the model. If a change to the code breaks the proof, the continuous verification rejects the code.
For unsafe code, we use RefinedRust to verify that functions adhere to Rust's safety contract for all inputs, ensuring that all clients only using safe Rust code cannot trigger memory unsafety.

\nip{Example: Page Tokens.}
As a simple but illustrative example of how our verification with {RefinedRust} works, we consider page tokens.
Page tokens are crucial to ensure memory isolation between security domains, \ie, no two security domains have access to the same page in confidential memory.
\autoref{fig:rr:pagetoken} shows excerpts of the invariant on page tokens,
 which specifies that page tokens are mathematically modeled by a type \rstinline{page} (defined in Rocq) with the following components: the memory location \rstinline{page_loc}, the stored sequence of words \rstinline{page_val}, and the size \rstinline{page_sz}.
Then, the core invariant is that a page token exclusively owns the memory region it spans, containing the list of words \rstinline{page_val}; for this, we provide a RefinedRust type assignment for the \rstinline{page_loc} location.
(Note that it is not possible to state this invariant directly in Rust's type system.)

The function \rstinline{read} on \rstinline{Page}s reads one machine word from the page
by using \emph{unsafe code} to read from the underlying memory.
Nevertheless, the function as a whole is \emph{safe} (\ie{} not marked as unsafe), as \rstinline{read} does appropriate bounds and alignment checks (\eg{}, that the read word is in confidential memory and inside the page) before performing the read, returning an error in case of an invalid input.
The function's specification defines the conditions under which the read succeeds and the expected result value.

The page token abstraction is useful as it safely abstracts from the low-level memory operations that it is implemented with.
With RefinedRust, we can verify once and for all that this abstraction is sound and upholds the safety guarantees of Rust's type system.
Other components of \ACE not using unsafe code (\eg{}, most of the page allocator) are thus automatically free of memory safety errors, even before fully verifying them for functional correctness and security -- and once we fully verify them, Rust's safety proof can be re-used to simplify the full security verification.

\begin{figure}
\begin{lstlisting}[language=rust]
#[rr::refined_by("p" : "page")]
/// Invariant: A Page exclusively owns its memory region.
#[rr::invariant(#type "p.(page_loc)" : "p.(page_val)" @
  "array usize (page_sz_in_words p.(page_sz))")]
/// Invariant: The page is well-formed.
#[rr::invariant("page_wf p")]
/// Omitted: the page resides in confidential memory
#[rr::invariant("...")]
pub struct Page {
    #[rr::field("p.(page_loc)")]
    address: ConfidentialMemoryAddress,
    #[rr::field("p.(page_sz)")]
    size: PageSize,
}
impl Page {
/// Precond: offset is divisible by the size of usize.
#[rr::requires("size_of usize | off_bytes")]
/// Precond: a usize fits at the offset within page bounds
#[rr::requires("off_bytes + size_of usize ≤
    page_sz_in_bytes p.(page_sz)")]
/// Postcond (omitted): return value is read from page
#[rr::ensures("...")]
fn read(&self, off_bytes: usize) -> Result<usize, Error> {
    unsafe { .. }
}}
\end{lstlisting}

\caption{Definition of the page token invariant in the RefinedRust specification language.}\label{fig:rr:pagetoken}
\end{figure}

\nip{Formal Verification Results and Bug Discovery.}
\cite{gaeher2026oopsla} shows how to formally verify memory safety and functional correctness (the two bottom layers of the pyramid above hardware in \autoref{fig:pyramide}) of the TSM's page management subsystem. The page token abstraction and the page allocator cover \textasciitilde600 lines of code, \textasciitilde8\% of the total TSM codebase.
The verification consists of 340 lines of expressive specifications written in RefinedRust, 1300 lines of shared Rocq specifications and 1400 lines of manually written Rocq proofs on top of the automatically-generated proofs.

The verification process uncovered two bugs in the unsafe parts of the page allocator implementation. The first bug was caused by incorrect bounds checking after pointer arithmetic in the procedure that creates the initial page tokens. The second was a logical error, identified during functional verification of the procedure responsible for assigning a new memory region to the page allocator.
Neither bug was detected during prior code review or testing, underscoring the value of formal methods for security-critical code.

Ongoing work focuses on establishing correspondence between the page table configuration and the RISC‑V authoritative semantics and integrating information‑flow tracking to enable non‑interference proofs with explicit reclassification to eventually reach the top of the pyramid in~\autoref{fig:pyramide}.
Spec and proof artifacts are part of the open source repository and verification is integrated with CI.
\section{Evaluation} \label{sec:eval}

\begin{figure*}[!t]
    \centering
    \includegraphics[width=\textwidth]{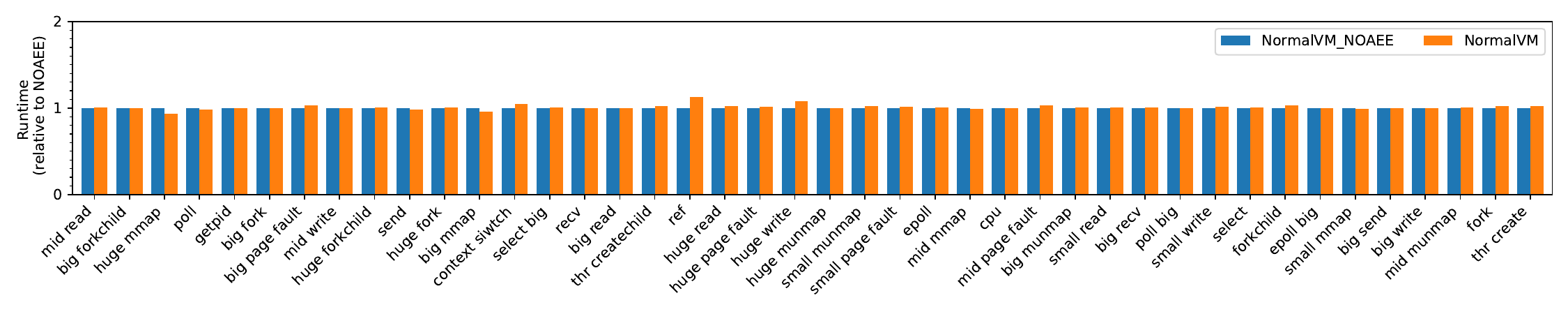}
    \caption{Overhead of running LEBench in a normal VM on system with \ACE to a system without \ACE. (higher=better)}
    \label{fig:lebench_noace}
\end{figure*}

\begin{figure*}[!t]
    \centering
    \includegraphics[width=\textwidth]{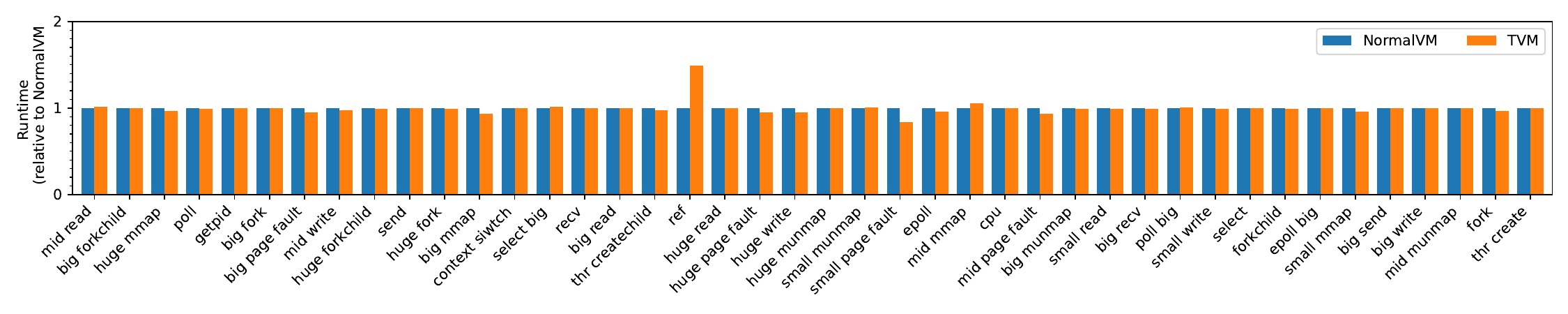}
    \caption{Overhead of running LEBench in a TVM vs a normal VM running on top of \ACE. (higher=better)}
    \label{fig:lebench_cvm}
\end{figure*}

We next evaluate whether \ACE's runtime performance and memory overhead of a TVM over a normal VM is similar to other confidential computing systems.
We seek to answer the following key questions:
\begin{itemize}
    \item What is the overhead of the TSM on TVM execution?
    \item How efficient is I/O in TVMs?
    \item What is the impact on a TVM's boot time?
    \item What is the implementation complexity and memory overhead of the page token mechanism?
\end{itemize}

\myparagraph{Evaluation setup.}
The evaluation hardware, which is the first commercially available RISC-V hardware supporting virtualization, implements a pre-ratified version of the hypervisor extension and does not fully conform to the CoVE specification. Specifically, three features require emulation in the TSM: VS-level timer virtualization, instruction fetch on trap, and clock access. These emulated paths introduce additional TSM execution overhead not present in CoVE-compliant hardware. We therefore treat our measurements as conservative upper bounds on \ACE's overhead: hardware conforming to CoVE will eliminate these emulation paths and reduce overhead.

Evaluation runs on a HiFive Premier P550 board equipped with an ESWIN EIC7700X SOC with four 64~bit RISC-V 1.4\,GHz imafdch cores.
The board has 16\,GB RAM, 8\,PMPs, and a Sv48x4 MMU. The host operating system runs Linux kernel 6.6.21 with CoVE and ESWIN patches that enable support for CoVE and P550 hardware. Since hardware does not implement a root-of-trust for attestation, we hard coded an attestation key to evaluate the local attestation mechanism.

\subsection{\ACE's Overhead}
To evaluate the CoVE patches' impact on the hypervisor and the TSM's runtime, we ran LEBench~\cite{lebench-paper} on VMs and TVMs with and without \ACE.
As the evaluation board's hardware clock does not support nanosecond accuracy, we modified LEBench to batch operations into a single timed measurement, ensuring a single measurement exceeds 1\,$\mu$s.

\autoref{fig:lebench_noace} shows that a normal VM executing on Linux KVM with CoVE patches and OpenSBI with the TSM performs mostly similarly to a normal VM running without a TSM.
With the TSM, the normal VM has lower performance for read/write, page fault, and context switch operations. This could be a result of additional execution paths and branches inside Linux KVM that were added by the experimental patch with support for the RISC-V nested acceleration extension (NACL) and a level of indirection introduced by the TSM. The CoVE spec requires NACL to share CSR values between the TSM and KVM and the current implementation affects both normal VMs and TVMs.

\autoref{fig:lebench_cvm} shows that core Linux kernel operations have similar performance for a TVM and a normal VM, except for mmap and page fault operations for which we observed between 5\% and 16\% performance increase for a TVM compared with a normal VM.
It might be caused by the way the TSM manages TVM's address translation, timers, and caches.
The TSM never pages out confidential memory pages, preventing G-stage page faults. Moreover, timer interrupt handling and scheduling happens directly in the TSM (M mode) while the regular VM's timer interrupts must be handled by KVM (HS mode) with the help of OpenSBI (M mode).

The ref benchmark, which measures kernel-managed clock reads, shows higher latency for TVMs. This is a consequence of the VS-level timer and clock access emulation in the TSM, both of which are absent from the evaluation hardware but mandated by the CoVE specification. Each clock read in a TVM triggers an additional trap through the TSM's emulation path, adding latency proportional to the TSM's context switch cost. This overhead is therefore a result of the evaluation platform and not a limitation of \ACE's design. CoVE-compliant hardware with Sstc support should eliminate this trap path.

\nip{\ACE's multi-vCPU overhead.} To get a better idea of whether the TSM causes any overhead when multiplexing between \mbox{vCPUs} within a VM, we use the parallel mode of CoreMark~\cite{coremark} to launch different numbers of instances of the benchmark inside a multi-VCPU VM.
Each VM is given 4, 8, and 16 vCPUs to match the benchmark's level of concurrency (number of parallel processes).
\autoref{tab:multivcpu} shows that the TSM does not introduce any significant performance differences when multiplexing the different vCPUs within a VM across the different configurations.

\begin{table}[b]
\centering
\caption{CoreMark benchmark results when running inside the respective VMs.
Reported values are means from three runs along with standard deviation.}
\label{tab:multivcpu}
\begin{tabular}{|l|c|c|c|}
\hline
\# of proc & VM\_ACE & VM & TVM \\ \hline
4 & 30003 $\pm$ 776 & 30387 $\pm$ 1172 & 29373 $\pm$ 713 \\ \hline
8 & 32191 $\pm$ 325 & 32368 $\pm$ 148 & 31614 $\pm$ 497\\ \hline
16 & 32952 $\pm$ 91 & 33055 $\pm$ 14 & 32822 $\pm$ 40 \\ \hline
\end{tabular}
\end{table}

\subsection{Network I/O}
\label{sec:eval:net}
We analyzed \ACE's impact on Virtio device performance by measuring network I/O overhead in a TVM. We measured throughput and latency of an Nginx server for each VM type and varying numbers of vCPUs.
The Nginx server and the ab benchmarking tool~\cite{abtool} run on separate machines connected to the same switch. ab used 8 threads to generate 10k requests for the same 615\,B file. VMs were configured with 5\,GB of memory and run Nginx with the number of workers matching the number of vCPUs. We ran each experiment 5 times and present the average with standard deviation.

\begin{figure}[!t]
    \centering
    \includegraphics[width=0.43\textwidth]{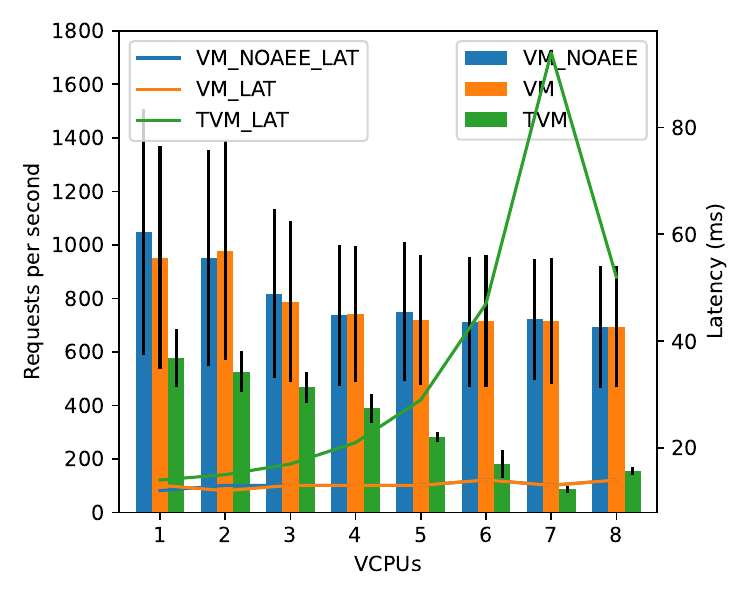}
    \caption{Network I/O performance when hosting Nginx server in VMs with different numbers of VCPUs.}
    \label{fig:nginx}
\end{figure}

\autoref{fig:nginx} shows that TVMs achieve 48\% of normal VM network throughput, the overhead consistent with the bounce-buffer-based DMA isolation common to VM-based confidential computing architectures (related work reports between  36\%~\cite{yan2023performance} and 60\%~\cite{misono2024confidential} performance overhead for network virtio operations on AMD SEV-SNP and Intel TDX).
This I/O overhead is architecturally inherent to any system that prevents the hypervisor and peripheral devices from directly accessing confidential memory resulting in an extra buffer copy for virtio.
On the evaluation board, we observe large throughput variation for normal VMs (between 550 to 1350 req/sec) and drop in throughput when adding more than 2 vCPUs. We attribute it to immature driver and firmware support on the evaluation board, which is the first RISC-V board on the market with early virtualization support.

There exist techniques to improve I/O performance in VM-based confidential computing system. One approach is to map a peripheral device (or a virtual function) directly into the TVM's address space, allowing the TVM to communicate with the device and enabling the device to read from and write to selected TVM pages without mediation by the hypervisor~\cite{tdisp,coveio}. An alternative approach avoids copying via bounce buffers by logically reassigning pages between confidential and non‑confidential memory using hardware-supported, page-based memory isolation~\cite{sahita2023smmtt}.
At present, no RISC‑V hardware supports these mechanisms.


\subsection{Boot Time}
We measured a VM boot time from the point we request its creation to the point we established an SSH connection to include overhead of local attestation, initialization of virtio devices, and network communication.
The results show an average time calculated over eight measurements.

\begin{figure}[!t]
    \centering
    \includegraphics[width=0.48\textwidth]{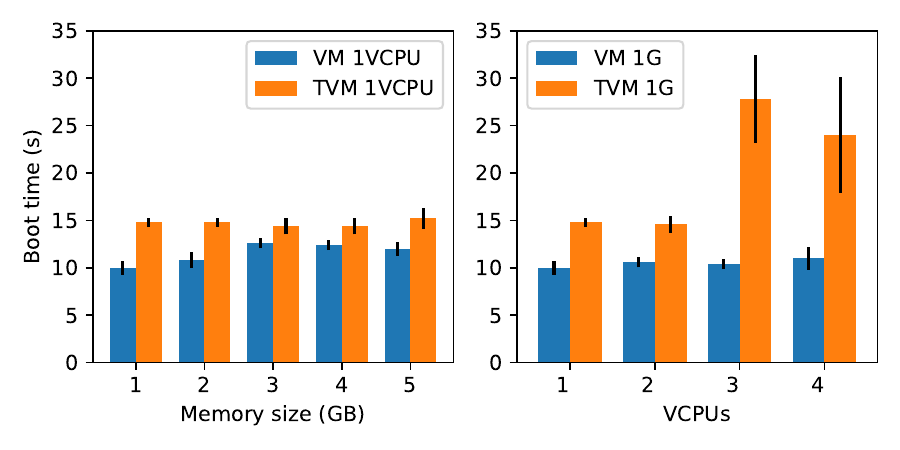}
    \caption{Avg VMs and TVMs boot time  with different memory size and vCPUs numbers. Bars show standard deviation.}
    \label{fig:boottime}
\end{figure}

\autoref{fig:boottime} presents boot times of normal VMs and TVMs for different memory sizes and numbers of vCPUs. Since a TVM's zeroed pages are lazily loaded, boot times for TVMs scale with the memory size.
TVMs boot up more slowly than normal VMs, 15\,sec vs 10\,sec.
This slower boot time is due to the Linux kernel creating thousands of shared pages needed for virtIO in an inefficient page-by-page approach, slower I/O caused by bounce buffers, and the local attestation check that performs signature check using a slow software implementation of ML-KEM cryptography.
The boot time latency can be thus improved with more efficient implementation (\ie, sharing multiple pages in a batch as allowed by the CoVE spec instead of \ACE's simpler implementation that does it page by page, hardware acceleration of cryptography operations) and more advanced mechanisms for improving I/O performance, as discussed in \S\ref{sec:eval:net}.
The increased boot time (up to 30\,sec) when adding more vCPUs to a TVM can be caused by the vCPUs synchronization: additional IPIs trigger more context switches resulting in cache misses due to security domain context switches.
\ACE does not impact a normal VM boot time compared to a non \ACE system.

\subsection{Page Token Memory Overhead}
A single page token occupies 9\,B in main memory regardless of the page size it represents. The page allocator stores page tokens in a tree data structure in which every level corresponds to a different architectural page size, adding extra 32\,B of information per not-empty node. The presence of a page token in the tree indicates that a page can be allocated.
To allow allocation of smaller pages, large page tokens are split into smaller ones.
An analogous operation, merging, occurs when pages are deallocated and returned to the page allocator. In such cases, multiple contiguous smaller page tokens are merged into a larger single page token, reducing the number of nodes in the tree. 
Thus, the latency of (de-)allocation operations is independent of fragmentation of the confidential memory, and the number of unallocated page tokens is minimized.
In more detail, 1\,GiB of unallocated confidential memory can be represented as a single 1\,GiB page token, taking 9\,B of the \ACE heap. If all page tokens representing the smallest architectural page size (4\,KiB) are allocated, then the page allocator's tree is empty but allocated page tokens are stored in the structures of a specific TVM. Then, the overhead of page tokens is the largest and results in occupancy of 2.25\,MiB of \ACE heap for every 1\,GiB of confidential memory (0.23\% of confidential memory).

Compared to other systems, Intel TDX and AMD SEV‑SNP require, in the pessimistic case, 16\,B of metadata per 4\,KiB physical page to track page assignment to confidential memory. This metadata is stored in hardware‑protected data structures known as the Physical Address Metadata Table (PAMT) and the Reverse Map Table (RMP), respectively. Overall, this corresponds to approximately 0.4\% of system memory~\cite{tdx_memoverhead, bagia2025close}.

\nip{Summary.}
Evaluation shows that \ACE is practical for our primary targets: compute-intensive, purpose-specific workloads such as automotive controllers, PCIe cryptographic accelerators, and telco infrastructure.
We expect next-generation CoVE-compliant hardware to reduce the I/O and boot overhead by eliminating the emulation paths described above.

\section{Related Work} \label{sec:related}

Commercial VM-based TEEs for x86~\cite{cheng2023intel,amd2020sev}, IBM Z~\cite{borntrager2020secure}, and OpenPOWER~\cite{hunt2021pef} primarily target high-end server processors in data centers.
In contrast, \ACE is tailored for edge and high-end embedded RISC-V systems, a target space largely unaddressed by existing confidential computing solutions. \ACE also introduces (1) a novel single-step TVM promotion mechanism that simplifies the ABI and eliminates intermediate attack surfaces~\cite{batteringramsp26,tee-fail}, (2) static memory partitioning coupled with formal verification as a way to build trust in the design and implementation, and (3) an attestation primitive for RISC-V systems with unreliable network access.
At the hardware level, the canonical design of \ACE runs on commodity RISC-V platforms while remaining extensible to use-case-specific hardware requirements.
As demonstrated by our evaluation, \ACE is deployable on commercially available silicon today.
By contrast, Salus, which also implements the RISC-V CoVE spec~\cite{cove_spec}, targets server-class environments and requires hardware extensions for scalable resource isolation (RISC-V Smmtt) and direct injection of VM interrupts (RISC-V AIA), limiting its current execution to simulation environments~\cite{sahita2023cove}.

ARM CCA~\cite{armcca} is a commercial VM-based confidential computing that targets application-class systems for ARM processors and requires dedicated hardware extensions.
It offers a more flexible architecture than \ACE that enables execution of multiple realm management monitor (RMM) instances, a firmware component analogous to \ACE's TSM.
ARM CCA also is subject to licensing fees.
In contrast, \ACE focuses on a simple but extensible  design that brings confidential computing capabilities to commodity RISC-V hardware.
To avoid adoption barriers in cost-sensitive regulated markets, \ACE is open-source and royalty-free.

Several academic and commercial systems target process-based confidential computing: Cerberus~\cite{LeeCerberus2022}, Keystone~\cite{lee2020keystone}, OP-TEE~\cite{OP-TEE}, Penglai~\cite{feng2021penglei}, Timber-V~\cite{weiser_2019_timberv}, Sanctum~\cite{costan2016sanctum}, SPEAR-V~\cite{SchrammelSPEAR-V2023}, and Elasticlave~\cite{Elasticlave}.
However, none of these fit our target systems because they do not scale above low-end to mid-range embedded systems due to limited architectural and hardware capabilities.
Example of target systems are smart meters, electronic control units, sensors, or single-purpose devices that must protect sensitive material, like cryptographic keys. In contrast, \ACE targets systems that support multiple security domains (e.g., multi-tenancy), run more complex workloads with custom multi-threading or IO requirements (\eg, automotive~\cite{redhat}, software defined vehicles, crypto and AI accelerators~\cite{nvidia_gpu}, telco~\cite{telco,vmware}).

Some of the confidential computing designs are subject to formal verification.
CertiKOS~\cite{certikos} is a framework for verified kernels that was used to verify the mCertiKOS hypervisor. It is written in an inherently unsafe combination of C and assembly, putting greater needs for abstraction on the verification methodology.
The verification proceeds in layers that progressively abstract implementation details in Rocq.
Li et al. \cite{cca_lietal} verify an early snapshot of the ARM CCA's RMM implementation using the CertiKOS approach, incrementally abstracting to a small top-level specification in Rocq.
\cite{cca_foxetal} propose continuous verification of the RMM using interactive theorem proving (HOL4) for a core model and bounded model checking and concurrency-aware testing for the actual system to achieve practical coverage at the cost of weaker assurances.
\ACE carries spec and proofs along the code and continuously verifies the changes in CI.
Verismo~\cite{verismo} functionally verifies firmware implemented in Rust and running inside a TVM on AMD SEV-SNP~\cite{amd2020sev}.
It relies on the isolation boundary provided by the AMD platform security processor (PSP), while \ACE's TSM is functionally closer to firmware running inside the PSP.
Their approach to formal verification differs from ours by using Verus~\cite{verus}: Verus is faster and more automated for proving functional specifications, while RefinedRust offers higher assurances due to its foundational verifier and smaller TCB.
Verus trusts unsafe Rust code, whereas we verify unsafe code.
Additionally, Verus's proofs require significant modifications to the Rust source code, while we verify idiomatic Rust code.

Komodo~\cite{ferraiuolo2017komodo}, ARM's process-based TEE, has limited potential to scale in terms of verification for VM-based TEE because of its assembly implementation.
Like \ACE, the NOVA~\cite{nova} microhypervisor minimizes the TCB by reducing to the core functionality. Ongoing work~\cite{bluerock_nova} is verifying a variant of NOVA.
GlobalPlatform~ \cite{GlobalPlatlformPPv1.3} publishes security standards for IoT devices through its Common Criteria Protection Profile (PP) for TEEs. ProveriT \cite{ProveriT} is a verified formal specification of the GlobalPlatform PP for TEEs. In contrast, we do not aim to just verify the specification, but the concrete implementation of \ACE.
\section{Conclusion}
\label{sec:conclud}
We have introduced principles and a methodology to design and implement high-assurance low-level systems.
We have applied these to build \ACE, an open-source and royalty-free confidential computing for edge and high-end embedded RISC-V systems. Thus far we have formally verified the memory safety of a core part of the \ACE implementation. The formal verification of the rest of the implementation and security properties is work in progress.
Our work on \ACE extends the RISC-V CoVE spec with a deployment model targeting edge and high-end embedded systems.

We provide the first evaluation of VM-based confidential computing on commercially available RISC-V hardware. The results show that a confidential VM running on this hardware (which does not
have confidential computing-specific extensions) is practical. \ACE incurs low performance overhead for process-intensive workloads and a competitive overhead up to 48\% for multi-vCPU network-intensive workloads.

\bibliographystyle{plain}
\bibliography{main}

\end{document}